\pdfoutput=1
\documentclass[a4paper]{article}
\usepackage[hmargin=1.5in]{geometry}

\usepackage{etoolbox}
\newcommand{\isarxiv}

\usepackage{glossaries}

\newacronym{rrt}{RRT}{rapidly-exploring random tree}
\newacronym{ocp}{OCP}{optimal control problem}
\newacronym{nlp}{NLP}{nonlinear program}
\newacronym{ned}{NED}{North East Down}
\newacronym{asv}{ASV}{autonomous surface vehicle}
\newacronym{colregs}{COLREGs}{International Regulations for Preventing Collisions at Sea}
\newacronym{prm}{PRM}{probabilistic roadmap}

\usepackage{algorithm}
\usepackage[noend]{algpseudocode}
	\algnewcommand{\algorithmbreak}{\textbf{break}}
	\algnewcommand{\Break}{\State \algorithmbreak}
	\algdef{SE}[DOWHILE]{Do}{doWhile}{\algorithmicdo}[1]{\algorithmicwhile\ #1}%
\usepackage{graphicx}
\usepackage{epstopdf}
\usepackage[output-product=\cdot,exponent-product=\cdot,per-mode=fraction,detect-weight=true]{siunitx}
\usepackage{nicefrac}
\usepackage{amsmath}
\usepackage{amsfonts}
\usepackage{amssymb}
\usepackage{mathtools}
\usepackage{bm}
\usepackage{commath}
\usepackage{booktabs}
\usepackage{tabulary}
\usepackage{placeins}
\usepackage{authblk}

\usepackage{natbib}
\usepackage[hyphens]{url}
\usepackage[hidelinks,breaklinks]{hyperref}

\newcommand{\astar}{\texorpdfstring{A$^\star$}{A*}}
\newcounter{step}
\newcommand{\step}[1]{\refstepcounter{step} \label{#1}}

\newcommand{\tr}{^\top}
\newcommand{\mtrx}[1]{\bm{\mathrm{#1}}}
\newcommand{\vect}[1]{\bm{#1}}
\newcommand{\mset}[1]{\mathbb{#1}}
\newcommand{\real}{\mathbb{R}}

\DeclareMathOperator{\scirc}{S}

\title{%
	Warm-Started Optimized Trajectory Planning for ASVs
}
\author{%
	Glenn Bitar,
	Vegard N.\ Vestad, \\
	Anastasios M.\ Lekkas and
	Morten Breivik
}

\date{%
	\footnotesize
	The authors are with the Centre for Autonomous Marine Operations and Systems, Department of Engineering Cybernetics, Norwegian University of Science and Technology (NTNU), NO-7491 Trondheim, Norway.
	E-mails: \{glenn.bitar,anastasios.lekkas\}@ntnu.no, vegardnittervestad@gmail.com, morten.breivik@ieee.org
	\\
	\vspace{0.5cm}
	\textcopyright{} 2019 IFAC
	}

\begin{document}
\maketitle
\begin{abstract}
	We consider warm-started optimized trajectory planning for autonomous surface vehicles (ASVs) by combining the advantages of two types of planners: an \astar{} implementation that quickly finds the shortest piecewise linear path, and an optimal control-based trajectory planner.
A nonlinear 3-degree-of-freedom underactuated model of an ASV is considered, along with an objective functional that promotes energy-efficient and readily observable maneuvers.
The \astar{} algorithm is guaranteed to find the shortest piecewise linear path to the goal position based on a uniformly decomposed map.
Dynamic information is constructed and added to the \astar{}-generated path, and provides an initial guess for warm starting the optimal control-based planner.
The run time for the optimal control planner is greatly reduced by this initial guess and outputs a dynamically feasible and locally optimal trajectory.

\end{abstract}

\FloatBarrier

\section{Introduction} 
\label{sec:introduction}

Motivated by potential for reduced costs, as well as safer and more environmentally friendly operations, technology for \glspl{asv} is being developed at a rapid pace.
Several commercial actors are spearheading the search for solutions for safe, collision-free and reliable autonomous operations.
Rolls-Royce and Finferries demonstrated the world's first autonomous car ferry ``Falco'' in 2018 \citep{Jallal2018FinferriesRR}, which navigated autonomously between two ports in Finland by combining advanced sensor technology and collision avoidance algorithms.

A prerequisite for safe and efficient operation is a well-functioning path or trajectory planning method.
Such a method is responsible for providing the \gls{asv} with a safe trajectory that avoids static obstacles such as land and shallow waters.
Depending on the type of operation, one might want to optimize the trajectory for various objectives, such as energy efficiency, speed or trajectory length.

\begin{figure}[tb]
	\centering
	\includegraphics[width=0.95\linewidth]{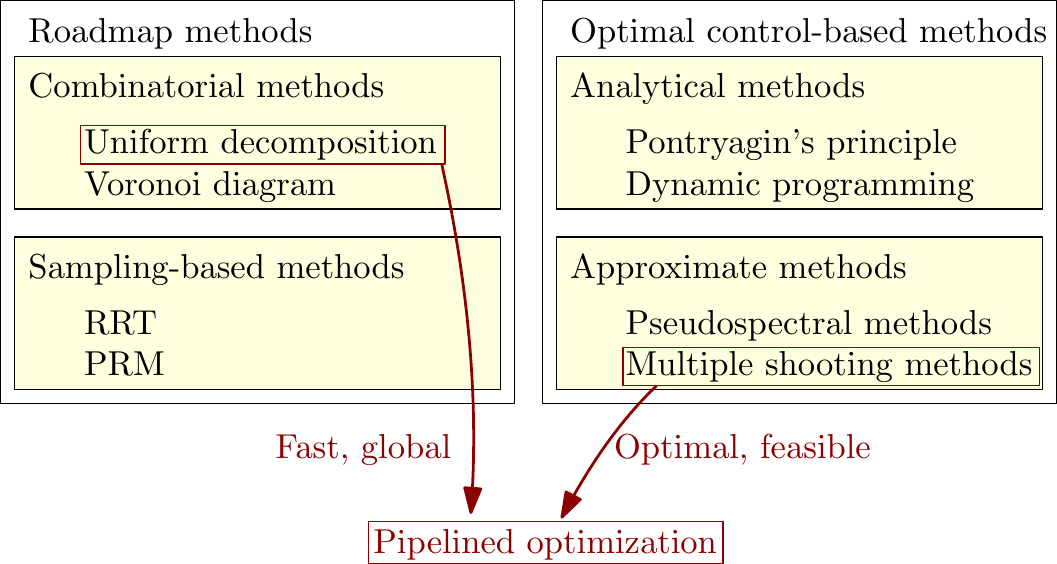}
	\caption{%
		Categorization of some planning algorithms.
	}
	\label{fig:planning-categorization}
\end{figure}

Numerous path and trajectory planning algorithms have been researched and are available for marine applications.
One may categorize these planning algorithms as being roadmap-based or optimization-based.
\autoref{fig:planning-categorization} gives an overview of the categorization of some planning algorithm types.
\emph{Roadmap methods} are based on exploring points in the geometric space in order to build a path between the start and goal positions.
There are two subcategories in roadmap methods.
\emph{Combinatorial} methods decompose an obstacle map using a preferred strategy, and perform a search in the resulting graph.
The decomposition strategies include e.g.\ uniform grids, Voronoi diagrams and visibility graphs.
The combinatorial methods explore the entire geometric space.
The graph search is often performed using \astar{}, which is an efficient and well-known search algorithm widely used to solve path planning problems \citep{Hart1968}.
\astar{} guarantees to find the shortest path when using an admissible heuristic function.
Hybrid \astar{} extends the \astar{} algorithm by generating dynamic trajectories to connect nodes, thus adding dynamic information to the search \citep{Dolgov2010}.
As opposed to combinatorial methods, \emph{sampling-based} methods randomly explores points in the map to build a path towards the goal.
\Gls{prm} is a sampling-based planning method that draws samples from the configuration space and connects them using a local planner \citep{Kavraki1996}.
A graph search algorithm is applied to find the minimum cost path from start to goal in the resulting graph.
\Gls{rrt} is another sampling-based method which calculates input trajectories between randomly sampled points and connects them in a tree until the start and goal positions are connected \citep{LaValle_1998}.
Although \gls{rrt} uses a cost function, the method is not optimal and will lock into the first connection between start and goal.
Various flavors of \gls{rrt} are developed to amend this, e.g.\ \gls{rrt}$^\star$ \citep{Karaman2011}.
This method continuously performs tree rewiring and has probabilistic completeness, but converges slowly.

The other group of planning methods contains algorithms based on \emph{optimal control}.
This group may further be divided into analytical and approximate methods.
Analytical methods such as Pontryagin's principle are only able to find solutions in very simple cases and are generally unpractical.
Approximate methods such as e.g.\ pseudospectral optimal control \citep{Bitar2018,Ross2012} are highly sensitive to initial guesses of the solution and will converge to a local optimum close to this guess.
Without a good initial guess, they also experience long run times and are sensitive to problem dimensionality.

\begin{figure}[tb]
	\centering
	\includegraphics[width=1.0\linewidth,height=5cm,keepaspectratio]{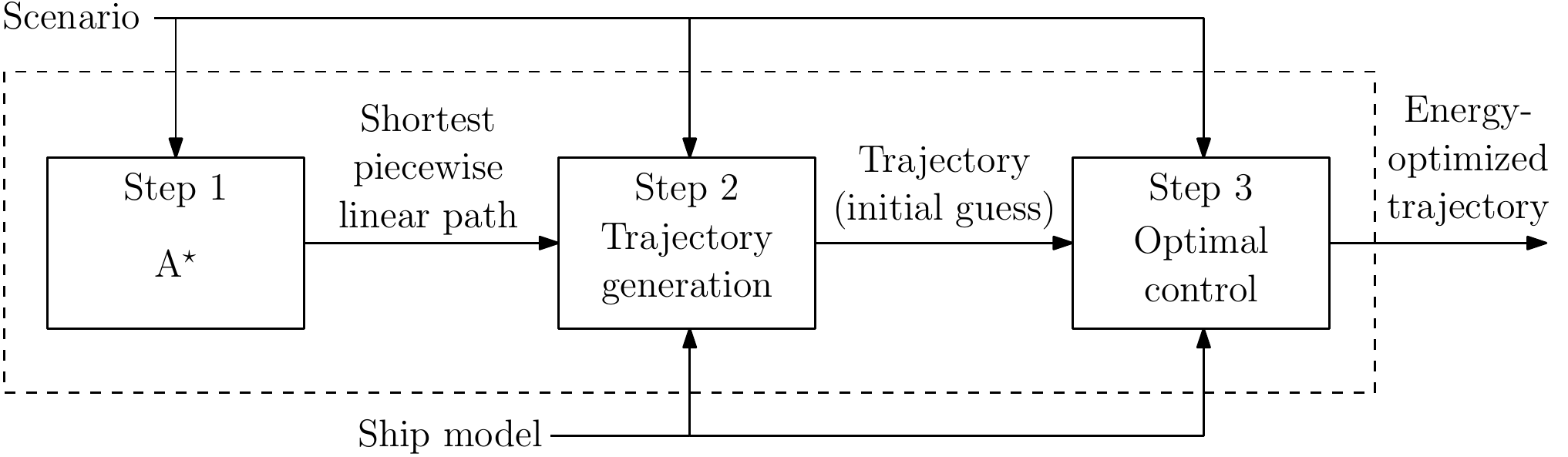}
	\caption{%
		Pipelined path planning concept.
	}
	\label{fig:concept-block-diagram}
\end{figure}

\citet{Zhang2018} plan trajectories for parking autonomous cars by combining hybrid \astar{} with an optimal control-based method.
Motivated by the same goals of exploiting the strengths and mitigate the weaknesses of optimal control-based algorithms, we here attempt to solve the long-term trajectory planning problem for \glspl{asv} as a transcribed \gls{ocp}, and warm start the solver using the smoothed solution of an \astar{} geometric planner.
In this three-step pipelined approach, the \astar{} planner swiftly provides a set of waypoints representing the shortest path as \autoref{step:astar}.
This path is converted into a full state trajectory by adding artificial and nearly feasible temporal information as \autoref{step:trajectory}.
\autoref{step:ocp} takes this trajectory and uses it as the initial guess for an \gls{ocp} solver, which finds an optimized trajectory near the globally shortest path.
The structure of this pipelined concept is illustrated in \autoref{fig:concept-block-diagram}.
The method is an off-line global planner, which assumes that information about the map and environment is known a priori.

The rest of this paper is organized as follows:
\autoref{sec:asv_modeling} presents the mathematical model of the \gls{asv} used in simulations and planning.
Finding the waypoints describing the shortest path with \astar{} is described in \autoref{sec:cell_decomposition}, and \autoref{sec:trajectory_generation} explains how the \astar{} solution is converted to a trajectory.
\autoref{sec:optimal_control} shows how the \gls{ocp} is transcribed to an \gls{nlp}, which yields an optimized trajectory when solved.
Simulation scenarios and results are presented in \autoref{sec:simulation_scenario_and_results}, while \autoref{sec:conclusion} concludes the paper.


\section{ASV modeling and obstacles} 
\label{sec:asv_modeling}

In \citep{Loe2008}, a simple nonlinear 3-degree-of-freedom ship model is identified to approximate the dynamics of the \gls{asv} Viknes 830.
Without loss of generality for the method described in this paper, we use that model to perform trajectory planning.
The model has the form
\begin{subequations}
	\label{eq:ship-model}
	\begin{align}
		\dot{\vect{\eta}} &= \mtrx{R}(\psi) \vect{\nu} \\
		\mtrx{M} \dot{\vect{\nu}} + \mtrx{C}(\vect{\nu}) \vect{\nu} + \mtrx{D}(\vect{\nu}) \vect{\nu} &= \vect{\tau}(\vect{u})\,.
	\end{align}
\end{subequations}
The pose vector $\vect{\eta} = [x, y, \psi]\tr \in \real^2 \times \scirc$ contains the \gls{asv}'s position and heading angle in the Earth-fixed \gls{ned} frame.
The velocity vector $\vect{\nu} = [u, v, r]\tr \in \mset{R}^3$ contains the \gls{asv}'s body-fixed velocities: surge, sway and yaw rate, respectively.
The rotation matrix $\mtrx{R}(\psi)$ transforms the body-fixed velocities to \gls{ned}:
\begin{equation}
	\mtrx{R}(\psi) =
	\begin{bmatrix}
		\cos \psi & -\sin \psi & 0 \\
		\sin \psi & \cos \psi & 0 \\
		0 & 0 & 1
	\end{bmatrix}.
\end{equation}
The matrix $\mtrx{M} \in \real^{3\times3}$ represents system inertia, $\mtrx{C}(\vect{\nu}) \in \real^{3\times3}$ Coriolis and centripetal effects, and $\mtrx{D}(\vect{\nu}) \in \real^{3\times3}$ represents damping effects.
The \gls{asv} is controlled by the control vector $\vect{u} = [X, N]\tr \in \real^2$, which contains surge force and yaw moment.
The control vector is mapped to a force vector $\vect{\tau}(\vect{u}) = [X, 0, N]\tr$.
The \gls{asv}'s states are collected in the vector $\vect{x} = [x, y, \psi, u, v, r]\tr$, and we collect the dynamic model \eqref{eq:ship-model} in the following compact form for notational ease in the remainder of the paper:
\begin{equation}
	\label{eq:ship-model-compact}
	\dot{\vect{x}} = \vect{f}(\vect{x}, \vect{u}) =
	\begin{bmatrix}
		\mtrx{R}(\psi) \vect{\nu} \\
		\mtrx{M}^{-1} \left( - \mtrx{C}(\vect{\nu}) \vect{\nu} - \mtrx{D}(\vect{\nu}) \vect{\nu} + \vect{\tau}(\vect{u}) \right)
	\end{bmatrix}.
\end{equation}


\section{Step 1: \astar{} path planner} 
\label{sec:cell_decomposition}
\step{step:astar}

To quickly find the global shortest collision-free path between a start and goal position, we use an \astar{} implementation on a uniformly decomposed grid.
The \astar{} implementation is standard, and details may be found in e.g.\ \citep{Hart1968}.
The search algorithm looks for collision-free paths between nodes in the uniform grid, and uses Euclidean distance as cost and heuristic functions.

The decomposition of the map affects the solution space and the run time for \autoref{step:astar}.
Using a uniform grid with grid size $\Delta d > 0$ too large will take paths going through narrow passages away from the solution space, and the desired shortest path may not be found.
A smaller grid size will explore more options, but requires more evaluation, giving a longer run time.
This uniform grid is in our case chosen for simplicity, however exploring other decompositions such as Voronoi diagrams or a non-uniform grid might be desirable for performance reasons.


\section{Step 2: Trajectory generation} 
\label{sec:trajectory_generation}
\step{step:trajectory}

In order to use the shortest path generated by \autoref{step:astar} as an initial guess for the \gls{ocp}, we convert it to a trajectory based on straight segments and circle arcs using a nominal forward velocity $u_{\text{nom}} > 0$.
The trajectory generation consists of three sub-steps: waypoint reduction, waypoint connection, and adding dynamic information.

\subsection{Waypoint reduction} 
\label{sub:waypoint_reduction}

\autoref{alg:waypoint-reduction} is employed to reduce the \astar{} path from \autoref{step:astar} to a minimum number of waypoints.
The algorithm outputs a reduced path as an ordered set of waypoints $\mset{P} = \set{\vect{p}_k \in \real^2~|~k=1,\dots,N_r}$ where $N_r$ is the number of waypoints.
The \astar{} waypoints are denoted $\vect{p}^\star_k$ for $k=1,\dots,N_\star$, ordered from start to goal, where $N_\star$ is the number of waypoints.

\begin{algorithm}[htb]
	\caption{Waypoint reduction algorithm.}
	\label{alg:waypoint-reduction}
	\begin{algorithmic}[1]
		\Procedure{Reduce}{}{}
			\State $i \gets N_\star$; $\mset{P} \gets \texttt{InitializePath}(\vect{p}^\star_i)$
			\Do
				\For{$j = 1$ to $i-1$}
					\If{$\lnot \texttt{Collision}(\vect{p}^\star_i, \vect{p}^\star_j)$}
						\State $\texttt{AddPoint}(\mset{P}, \vect{p}^\star_j)$
						\State $i \gets j$
						\Break
					\EndIf
				\EndFor
			\doWhile{ $i > 1$ }
		\EndProcedure
	\end{algorithmic}
\end{algorithm}


\subsection{Waypoint connection} 
\label{sub:waypoint_connection}

The waypoints in the reduced path $\vect{p}_k \in \mset{P}$ are connected with straight segments and circle arcs to increase geometric feasibility.
This is done by calculating the parameters of a circle based on a radius of acceptance $R_{\text{acc}} > 0$.
The result is a path with discontinuous turn rate since the turn rate of such a curve will experience jumps at the beginning and end of the circle arcs.
However, if the circle arcs have a turning radius $R_{\text{turn}} > 0$ larger than the minimum turning radius of the \gls{asv} $R_{\text{turn},\text{min}} > 0$, the resulting geometry of the path can be followed tightly.
Additional information about such a path waypoint connection is available in \citep{Fossen2011}.

For each straight segment, the turn rate is $0$.
For the circle arcs, the turn rate is $\nicefrac{u_{\text{nom}}}{R_{\text{turn},k}}$, where $R_{\text{turn},k}>0$ is the turning radius for arc $k$.
The tangent angles for the straight segments are $\gamma_k \in \scirc$, and for the circle arcs, the tangent angles move between $\gamma_k$ and $\gamma_{k+1}$, depending on how far along the curve it is evaluated.

Using this information, we can concatenate a path consisting of alternations of straights and circle arcs, and construct a path function parametrized by length with position
\begin{subequations}
\begin{equation}
	\vect{p}_g: [0, L_{\text{path}}] \to \real^2\,,
\end{equation}
where $L_{\text{path}} > 0$ is the total length of the path.
Functions for path tangential angle and turn rate are also constructed:
\begin{align}
	\gamma_g&: [0, L_{\text{path}}] \to \scirc \text{, and}\\
	r_g&: [0, L_{\text{path}}] \to \real\,,
\end{align}
\end{subequations}
respectively.
These functions are subscripted by $(\cdot)_g$ to indicate that they are based on the path geometry.


\subsection{Adding temporal information} 
\label{sub:adding_dynamic_information}

After obtaining an arc-length parametrized path we add temporal information by assuming a constant surge velocity $u_{\text{nom}}$, a sway velocity $v$ of zero, and piecewise constant yaw rate $r$.
The nominal surge velocity is determined by $u_{\text{nom}} = \frac{L_{\text{path}}}{t_{\text{max}}}$, where $t_{\text{max}} > 0$ is the tunable time to complete the trajectory, which is valid on $t \in [0, t_{\text{max}}]$.
The distance traveled will be $L(t) = u_{\text{nom}} \cdot t$, and the states will then have trajectories
\begin{subequations}
\begin{align}
	\begin{bmatrix}
		x_w(t) & y_w(t)
	\end{bmatrix}\tr
	&= \vect{p}_g(L(t)) \\
	\psi_w(t) &= \gamma_g(L(t)) \\
	u_w(t) &= u_{\text{nom}} \\
	v_w(t) &= 0 \\
	r_w(t) &= r_g(L(t)) \,.
\end{align}
\end{subequations}
The input trajectory is set to the constant values
\begin{equation}
	\tau_{X,w}(t) = \tau_{X,ss}, \quad \tau_{N,w}(t) = 0\,
\end{equation}
where $\tau_{X,ss} \in \real$ is calculated as the steady-state value required to maintain nominal forward velocity $u_{\text{nom}}$.
The trajectories are subscripted by $(\cdot)_w$ to indicate that they will be used for warm-starting the \gls{ocp} in \autoref{step:ocp}.

The resulting trajectory is not dynamically feasible according to \eqref{eq:ship-model} but will be used as an initial guess for the \gls{ocp} solver, described in the next section.
The trajectory is collected in the following vectors:
\begin{equation}
	\vect{x}_w(t) =
	\begin{bmatrix}
		x_w(t) \\
		y_w(t) \\
		\psi_w(t) \\
		u_w(t) \\
		v_w(t) \\
		r_w(t)
	\end{bmatrix}
	\ 
	\vect{u}_w(t) =
	\begin{bmatrix}
		\tau_{X,w}(t) \\
		\tau_{N,w}(t)
	\end{bmatrix}\ 
	\forall \, t \in [0, t_{\text{max}}]\,.
\end{equation}

The goal of the method described in this paper is to find a trajectory of states and inputs that minimizes a cost functional $J(\vect{x}(\cdot), \vect{u}(\cdot))$:
\begin{equation}
	\label{eq:cost-functional}
	J(\vect{x}(\cdot), \vect{u}(\cdot)) =
		\int_{0}^{t_{\text{max}}}
			F(\vect{x}(\tau), \vect{u}(\tau))
		\dif \tau\,,
\end{equation}
which is dependent on a cost-to-go function $F(\vect{x}, \vect{u})$.
This function may be selected to find e.g.\ the trajectory that minimizes energy usage.
The initial guess for the cost trajectory $J_w(\cdot)$ at time $t$ is determined by
\begin{equation}
	\label{eq:cost-to-go-integral}
	J_w(t) = \int_0^t F(\vect{x}_w(\tau), \vect{u}_w(\tau)) \dif \tau\,.
\end{equation}



\section{Step 3: Optimal control} 
\label{sec:optimal_control}
\step{step:ocp}

Optimal control is used to make feasible and optimize the trajectory provided by \autoref{step:trajectory}.
An \gls{ocp} is formulated as
\begin{subequations}
\label{eq:ocp}
\begin{align}
	\label{eq:ocp-cost}
	&\min_{\vect{x}(\cdot), \vect{u}(\cdot)} \int_{0}^{t_{\text{max}}} F(\vect{x}(\tau), \vect{u}(\tau)) \dif \tau \\
	\nonumber
	&\text{subject to} \\
	\label{eq:ocp-dynamics}
	&\dot{\vect{x}}(t) = \vect{f}(\vect{x}(t), \vect{u}(t)) ~ \forall t \in [0, t_{\text{max}}] \\
	\label{eq:ocp-ineq}
	&\vect{h}(\vect{x}(t), \vect{u}(t)) \leq \vect{0} ~ \forall t \in [0, t_{\text{max}}] \\
	\label{eq:ocp-boundary}
	&\vect{e}(\vect{x}(0), \vect{x}(t_{\text{max}})) = \vect{0}\,.
\end{align}
\end{subequations}
The solution of this \gls{ocp} gives a trajectory of states $\vect{x}(\cdot)$ and inputs $\vect{u}(\cdot)$ that minimizes \eqref{eq:cost-functional}.

\subsection{Cost functional} 
\label{sub:cost_functional}

The cost functional described in \eqref{eq:cost-functional} is dependent on the cost-to-go function $F(\vect{x}, \vect{u})$.
This function may be adjusted and structured according to the desired sense of optimality.
Our aim is a trajectory which is optimized for energy usage, as well as performing readily observable maneuvers, as is required by \gls{colregs} Rule~8.
This results in a two-part cost-to-go function:
\begin{equation}
	\label{eq:cost-to-go}
	F(\vect{x}, \vect{u}) = K_e F_e(\vect{x}, \vect{u}) + K_t F_t(\vect{x})\,,
\end{equation}
with tuning parameters $K_e, K_t > 0$.
The first term penalizes energy usage and describes work done by the actuators:
\begin{equation}
	\label{eq:cost-energy}
	F_e(\vect{x}, \vect{u}) = \abs{u\cdot\tau_X} + \abs{r\cdot\tau_N}\,.
\end{equation}
The second term is a disproportionate penalization on turn-rate $r$, which prefers readily observable turns performed with high turn-rate.
The function has the form
\begin{equation}
	\label{eq:cost-turn}
	F_t(\vect{x}) = \left(a_t r^2 + (1 - e^{-\frac{r^2}{b_t}})\right) \frac{1}{F_{t,max}}\,,
\end{equation}
where
\begin{equation}
	F_{t,max} = a_t r_{\text{max}}^2 + (1 - e^{-\frac{r_{\text{max}}^2}{b_t}})\,,
\end{equation}
and $r_{\text{max}} > 0$ is the \gls{asv}'s maximum yaw rate.
The tuning parameters $a_t > 0$ and $b_t > 0$ shape the penalization to prefer higher or lower turn-rate, which is an idea obtained from \citep{Eriksen2017MPC}.


\subsection{Obstacles} 
\label{sub:obstacles}

Obstacles are encoded as elliptic inequalities in \eqref{eq:ocp-ineq}.
The basis for one elliptic obstacle is
\begin{equation}
	\left(\frac{x-x_c}{x_a}\right)^2 + \left(\frac{y-y_c}{y_a}\right)^2 \geq 1\,,
\end{equation}
where $x_c$ and $y_c$ describe the ellipse center and $x_a$ and $y_a$ describe the sizes of the two elliptic axes.
The ellipses are rotated by $\alpha$, which is the angle between the global $x$-axis and the direction of $x_a$.
The resulting inequality becomes
\begin{multline}
	\label{eq:ellipse}
		g_{o}(x, y, x_c, y_c, x_a, y_a, \alpha) = \\
		- \log \bigg[
		\left( \frac{ (x - x_c) \cos \alpha + (y - y_c) \sin \alpha}{x_a} \right)^2 \\
		+ \left( \frac{-(x - x_c) \sin \alpha + (y - y_c) \cos \alpha}{y_a} \right)^2 + \epsilon \bigg] \\
		+ \log(1 + \epsilon) \leq 0\,,
\end{multline}
where a small value $\varepsilon>0$ is added to deal with feasibility issues as $x \to x_c$ and $y \to y_c$, and the logarithmic function is used to reduce numerical sizes, without changing the inequality.
The same function is used in \citep{Bitar2019hybridcolav}.


\subsection{NLP transcription} 
\label{sub:nlp_transcription}

A multiple-shooting approach is used to transcribe the \gls{ocp} into an \gls{nlp}:
\begin{subequations}
\label{eq:nlp}
\begin{align}
	\label{eq:nlp-cost}
	&\min_{\vect{w}} \phi(\vect{w}) \\
	\nonumber
	&\text{subject to} \\
	\label{eq:nlp-ineq}
	& \vect{g}_{lb} \leq \vect{g}(\vect{w}) \leq \vect{g}_{ub} \\
	\label{eq:nlp-bounds}
	&\vect{w}_{lb} \leq \vect{w} \leq \vect{w}_{ub} \,.
\end{align}
\end{subequations}

The dynamics are discretized into $N_{ocp}$ steps in time, with step length $h=\nicefrac{t_{\text{max}}}{N_{ocp}}$.
The decision variables $\vect{w}$ consist of the state variables $\vect{x}_k = \vect{x}(t_k),\, k = 0,1,\dots,N_{ocp}$, the accumulated costs $J_k = J(t_k),\, k = 0,1,\dots,N_{ocp}$, where
\begin{equation}
	\label{eq:cost-integral}
	J(t) = \int_0^t F(\vect{x}(\tau), \vect{u}(\tau)) \dif \tau\,,
\end{equation}
and the control inputs $\vect{u}_k = \vect{u}(t_k),\,k=0,1,\dots,N_{ocp}-1$:
\begin{equation}
	\label{eq:decision-variables}
	\vect{w} =
	\begin{bmatrix}
		\vect{z}_0\tr & \vect{u}_0\tr & \vect{z}_1\tr & \dots & \vect{u}_{N_{ocp}-1}\tr & \vect{z}_{N_{ocp}}\tr
	\end{bmatrix}\tr,
\end{equation}
where $\vect{z}_k = [\vect{x}_k\tr, J_k]\tr$.

The cost function \eqref{eq:nlp-cost} approximates \eqref{eq:ocp-cost} and is
\begin{equation}
	\phi(\vect{w}) = J_{N_{ocp}}\,.
\end{equation}

The constraints \eqref{eq:nlp-ineq} are used to satisfy shooting constraints, as well as the collision avoidance constraints.
For the shooting constraints, we construct a discrete representation of the dynamics \eqref{eq:ocp-dynamics} as well as the integral \eqref{eq:cost-integral} using a RK4 scheme with $K_{ocp}$ steps.
We define the discrete version of \eqref{eq:ocp-dynamics} augmented with the time derivative of \eqref{eq:cost-integral} as
\begin{equation}
	\vect{z}_{k+1}
	= \vect{F}(\vect{z}_k, \vect{u}_k)\,,
\end{equation}
and construct the shooting constraints
\begin{equation}
	\vect{g}_{s}(\vect{w}) =
	\begin{bmatrix}
		\vect{z}_1 - \vect{F}(\vect{z}_0, \vect{u}_0) \\
		\vdots \\
		\vect{z}_{N_{ocp}} - \vect{F}(\vect{z}_{N_{ocp}-1}, \vect{u}_{N_{ocp}-1})
	\end{bmatrix},
\end{equation}
with associated lower and upper bounds
\begin{equation}
	\vect{g}_{s,lb} = \vect{g}_{s,ub} = \vect{0}_{(n+1)\cdot N_{ocp}}\,.
\end{equation}

For obstacles $i=1,2,\dots,N_o$, we avoid collisions by satisfying the inequality constraint
\begin{equation}
	g_o(x_k, y_k, x_{c,i}, y_{c,i}, a_i, b_i, \alpha_i) \leq 0\,,
\end{equation}
where $x_k = x(t_k)$ and $y_k = y(t_k)$ for $k = 1,2,\dots,N_{ocp}$.
We create a vector for all our obstacles in a single time step:
\begin{equation}
	\begin{gathered}
		\bar{\vect{g}}_o(\vect{x}_k) = \\
		\begin{bmatrix}
			g_o(x_k, y_k, x_{c,1}, y_{c,1}, a_{1}, b_{1}, \alpha_1) \\
			g_o(x_k, y_k, x_{c,2}, y_{c,2}, a_{2}, b_{2}, \alpha_2) \\
			\vdots \\
			g_o(x_k, y_k, x_{c,N_o}, y_{c,N_o}, a_{N_o}, b_{N_o}, \alpha_{N_o})
		\end{bmatrix}.
	\end{gathered}
\end{equation}
Obstacle constraints for all time steps are gathered in
\begin{equation}
	\vect{g}_o(\vect{w}) =
	\begin{bmatrix}
		\bar{\vect{g}}_o(\vect{x}_0) \\
		\bar{\vect{g}}_o(\vect{x}_1) \\
		\vdots \\
		\bar{\vect{g}}_o(\vect{x}_{N_{ocp}-1}) \\
	\end{bmatrix}
\end{equation}
with associated lower and upper bounds
\begin{equation}
	\vect{g}_{o,lb} = - \vect{\infty}_{N_{o}\cdot N_{ocp}} \quad \text{and} \quad \vect{g}_{o,ub} = \vect{0}_{N_{o}\cdot N_{ocp}}\,.
\end{equation}

The nonlinear inequality constraints \eqref{eq:nlp-ineq} are completed as
\begin{equation}
	\vect{g}_{lb} =
	\begin{bmatrix}
		\vect{g}_{s,lb} \\
		\vect{g}_{o,lb}
	\end{bmatrix},~
	\vect{g}(\vect{w}) =
	\begin{bmatrix}
		\vect{g}_s(\vect{w}) \\
		\vect{g}_o(\vect{w})
	\end{bmatrix},~
	\vect{g}_{ub} =
	\begin{bmatrix}
		\vect{g}_{s,ub} \\
		\vect{g}_{o,ub}
	\end{bmatrix}.
\end{equation}

The decision variable bounds \eqref{eq:nlp-bounds} are used to satisfy constant state and input constraints, as well as boundary conditions \eqref{eq:ocp-boundary}.
The bounds are
\begin{subequations}
\begin{align}
	\begin{gathered}
		\vect{w}_{lb}\tr = \\
		\begin{bmatrix}
			\vect{z}_{s,lb}\tr & \vect{u}_{lb}\tr & \vect{z}_{lb}\tr & \vect{u}_{lb}\tr & \dots & \vect{u}_{lb}\tr & \vect{z}_{f,lb}\tr
		\end{bmatrix}
	\end{gathered} \\
	\begin{gathered}
		\vect{w}_{ub}\tr = \\
		\begin{bmatrix}
			\vect{z}_{s,ub}\tr & \vect{u}_{ub}\tr & \vect{z}_{ub}\tr & \vect{u}_{ub}\tr & \dots & \vect{u}_{ub}\tr & \vect{z}_{f,ub}\tr
		\end{bmatrix},
	\end{gathered}
\end{align}
\end{subequations}
where
\begin{subequations}
\begin{align}
	\vect{z}_{s,lb} &=
	\begin{bmatrix}
		x_s & y_s & \psi_{lb} & u_{r,s} & 0 & 0 & 0
	\end{bmatrix}\tr \\
	\vect{z}_{s,ub} &=
	\begin{bmatrix}
		x_s & y_s & \psi_{ub} & u_{r,s} & 0 & 0 & 0
	\end{bmatrix}\tr \\
	\vect{z}_{f,lb} &=
	\begin{bmatrix}
		x_f & y_f & \psi_{lb} & u_{r,lb} & 0 & 0 & 0
	\end{bmatrix}\tr \\
	\vect{z}_{f,ub} &=
	\begin{bmatrix}
		x_f & y_f & \psi_{ub} & u_{r,ub} & 0 & 0 & \infty
	\end{bmatrix}\tr \\
	\vect{z}_{lb} &=
	\begin{bmatrix}
		x_{lb} & y_{lb} & \psi_{lb} & u_{r,lb} & v_{lb} & r_{lb} & 0
	\end{bmatrix}\tr \\
	\vect{z}_{ub} &=
	\begin{bmatrix}
		x_{ub} & y_{ub} & \psi_{ub} & u_{r,ub} & v_{ub} & r_{ub} & \infty
	\end{bmatrix}\tr \\
	\vect{u}_{lb} &=
	\begin{bmatrix}
		X_{lb} & N_{lb}
	\end{bmatrix}\tr \\
	\vect{u}_{ub} &=
	\begin{bmatrix}
		X_{ub} & N_{ub}
	\end{bmatrix}\tr,
\end{align}
\end{subequations}
and where values subscripted with $(\cdot)_s$ represent initial conditions, $(\cdot)_f$ the final conditions, and $(\cdot)_{lb}$ and $(\cdot)_{ub}$ represent lower and upper bounds, respectively.


\subsection{Initial guess and solver} 
\label{sub:initial_guess}

The trajectories $\vect{x}_w(\cdot)$, $\vect{u}_w(\cdot)$ and $J_w(\cdot)$ from \autoref{sub:initial_guess} are used as an initial guess to warm-start the \gls{nlp}.
These trajectories are sampled at the time steps $t_k$, $k=0,\dots,N_{ocp}$ using interpolation, and shaped into the form of the decision vector $\vect{w}$ \eqref{eq:decision-variables}, providing the initial guess $\vect{w}_0$.


The \gls{nlp} as defined by \eqref{eq:nlp} is solved by the interior-point method Ipopt \citep{Wachter2005ipopt} using Casadi \citep{Andersson2018casadi} for Matlab.

\subsection{Algorithm summary} 
\label{sub:algorithm_summary}

The pipelined algorithm is summarized by the steps in \autoref{tab:algorithm-steps}, where the properties of each step are highlighted in terms of parametrization, feasibility according to \eqref{eq:ship-model} and optimality.

\begin{table}[tb]
	\centering
	\caption{Algorithm step explanation.}
	\begin{tabulary}{1.0\linewidth}{CCCC}
		\toprule
		Step					& Parametrized by	& Dynamic feasibility					&  Optimality \\
		\midrule
		\ref{step:astar}		& Length			& None, piecewise linear				&  Shortest piecewise linear path \\
		\ref{step:trajectory}	& Time				& Discontinuous yaw rate $r$			&  None \\
		\ref{step:ocp}			& Time				& Adheres to \eqref{eq:ship-model}		&  Energy and \acrshort{colregs} Rule~8 \\
		\bottomrule
	\end{tabulary}
	\label{tab:algorithm-steps}
\end{table}

While \autoref{step:astar} gives the shortest piecewise linear path, it is parametrized by length, and will not be dynamically feasible for warm-starting the \gls{ocp} in \autoref{step:ocp}.
\autoref{step:trajectory} connects the waypoints with circle arcs and adds artificial dynamics, which moves us closer to a dynamically feasible trajectory.
However, we lose the optimality of the shortest path with this modification, and the yaw rate is discontinuous, which is not possible according to \eqref{eq:ship-model}.
This trajectory is usable as an initial guess for \autoref{step:ocp}, which converges to a trajectory that adheres to \eqref{eq:ship-model}, and adds optimality according to \eqref{eq:ocp-cost}.



\section{Simulation scenarios and results} 
\label{sec:simulation_scenario_and_results}

\begin{figure}[tb]
	\centering
	\includegraphics[width=0.95\linewidth,height=0.5\textheight,keepaspectratio]{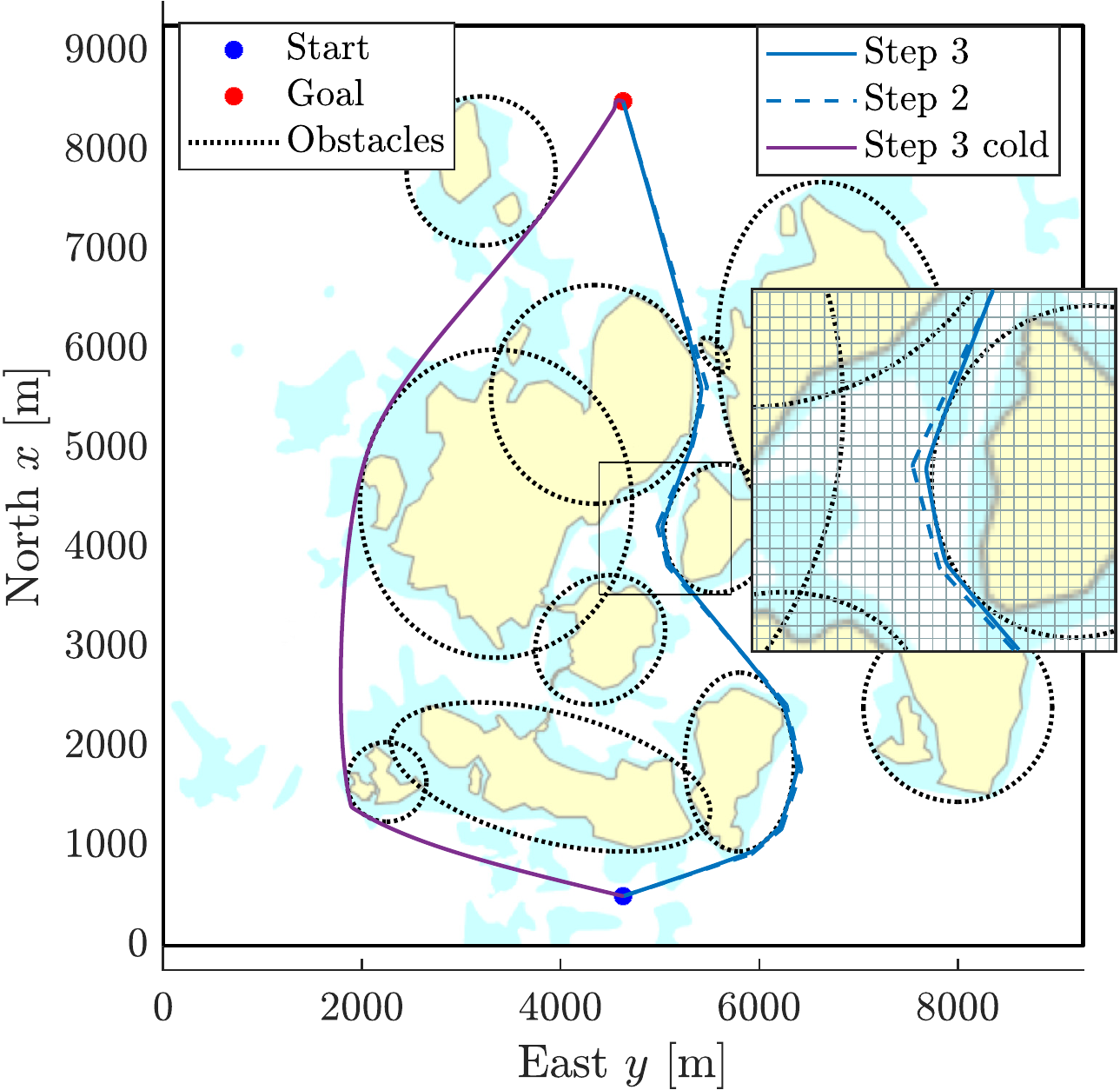}
	\caption{%
		Map showing the scenario used for planning, with multiple elliptical obstacle boundaries surrounding the small islands.
		Trajectories after steps~\ref{step:trajectory} and~\ref{step:ocp} are plotted.
		A cold-started solution is also included.
	}
	\label{fig:overlaid-paths}
\end{figure}

The scenario selected for testing our planning method is Sjernar{\o}y north of Stavanger, Norway, near \SI{59.25}{\degree}N and \SI{5.83}{\degree}E.
A map of this scenario is shown in \autoref{fig:overlaid-paths}.
The scenario has many possible routes between the start and goal positions, including routes that go outside the islands, and the narrow passage between the islands.
The narrow passage is the shortest path, and one could claim that in the absence of disturbances, this shortest path is also the most energy efficient.
However, since the problem of finding this path is non-convex and resembles an integer problem, the \gls{ocp} alone would struggle to find the shortest path.
We use the algorithm parameters presented in \autoref{tab:parameter-values}.

\begin{table}[tb]
	\centering
	\caption{Parameter values.}
	\begin{tabular}{cc|cc}
		\toprule
		Param. 				& Val.  											& Param. 						& Val. \\
		\midrule
		$\Delta d$ 			& \SI{50}{[\meter]}									& $t_{\text{max}}$ 				& \SI{2200}{[\second]}									\\
		$N_{ocp}$ 			& \SI{1000}{}								& $K_{ocp}$ 					& \SI{1}{}										\\
		$K_e$				& \SI{3.5e-4}{[\joule^{-1}]} 						& $K_t$							& \SI{800}{} 									\\
		$a_t$ 				& \SI{112}{[\second\squared\per\radian\squared]}	& $b_t$							& \SI{6.25e-5}{[\radian\squared\per\second\squared]}	\\
		$R_{acc}$ 			& \SI{10}{[\meter]}									& $R_{\text{turn},\text{min}}$ 	& \SI{24.5}{[\meter]}									\\
		$r_{\text{max}}$	& \SI{40}{[\degree\per\second]} \\
		\bottomrule
	\end{tabular}
	\label{tab:parameter-values}
\end{table}

To benchmark our planning algorithm, we apply it to the scenario illustrated in \autoref{fig:overlaid-paths} in Matlab on a laptop with an Intel Core i7-7700HQ processor.
For comparison, we also apply the \gls{ocp} to the same scenario without an initial guess, i.e.\ cold starting \autoref{step:ocp}.
Solutions from these two methods will be dynamically feasible trajectories with different routings to reach the goal position.
We use metrics of total cost and run times to compare the algorithms.
These metrics will also be applied to the trajectory after \autoref{step:trajectory}.
This trajectory is not dynamically feasible according to \eqref{eq:ship-model} but can tell us how the smoothed \astar{} trajectory performs without optimization.

The resulting trajectories are plotted on top of the scenario in \autoref{fig:overlaid-paths}.
We see that the initial guess goes through the narrow passage between the islands and that the warm-started \gls{ocp} finds a solution along the same route.
As expected, the cold-started \gls{ocp} goes outside the passage and finds a longer solution.
A zoomed inset in \autoref{fig:overlaid-paths} shows how the \gls{ocp} is able to produce readily observable maneuvers by making sharp turns around the obstacle boundaries.
The inset also includes the grid used by \autoref{step:astar}.

\begin{figure}[tb]
	\centering
	\includegraphics[width=0.95\linewidth]{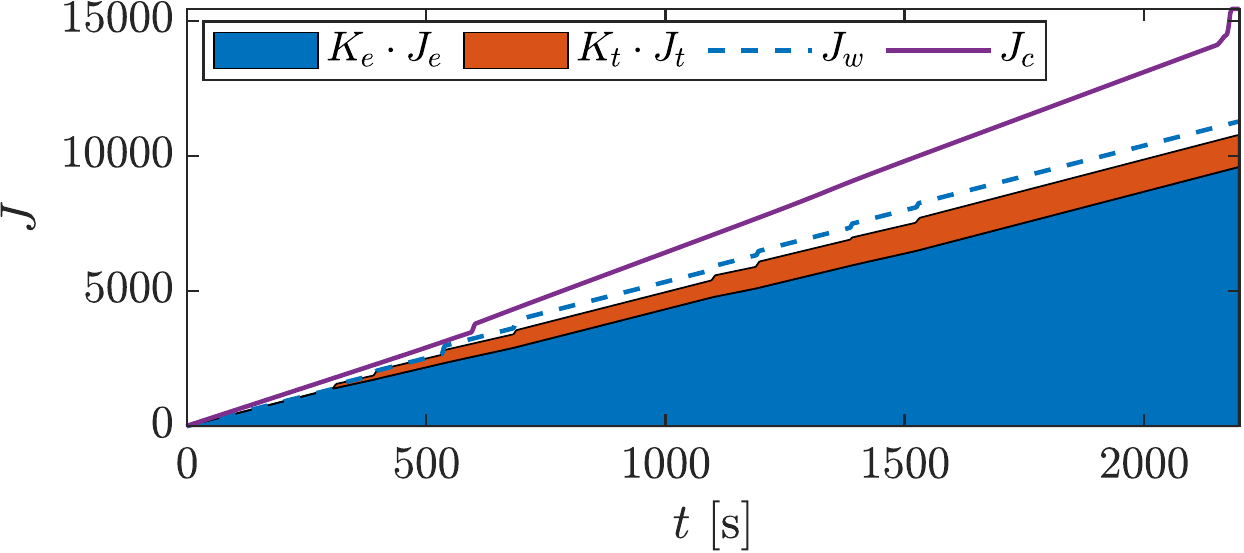}
	\caption{%
		Cost functional development along both the optimized trajectory and the initial guess.
		The optimized trajectory shows the cost split into contributions from energy optimization and observable maneuvers.
		Also, the cost of the cold-started \acrshort{ocp} is denoted $J_c$.
	}
	\label{fig:cost}
\end{figure}

\autoref{fig:cost} shows us the cost functional develops along the trajectories of the warm-started \gls{ocp} ($K_e \cdot J_e + K_t \cdot J_t$), the initial guess ($J_w$) and the cold-started \gls{ocp} ($J_c$).
\autoref{tab:results} shows the results at $t=t_{\text{max}}$ for the three methods.
We see the scaled total cost as calculated by \eqref{eq:ocp-cost} and \eqref{eq:cost-to-go}, as well as the energy cost calculated by \eqref{eq:cost-energy}.
An improvement of \SI{30}{\percent} is obtained by warm-starting the \gls{ocp} compared to cold starting it, explained by the shorter route selection.
The warm-started \gls{ocp} is also able to improve on the dynamically infeasible initial guess by \SI{4}{\percent}.

\autoref{tab:results} also shows the run times of the three methods.
Since the initial guess alone does not perform any iterative optimization, it has the lowest run time.
The warm-started method spends \SI{27}{\second} in total to find an optimized solution to the path planning problem, including \SI{21}{\second} spent solving the \gls{ocp}.
This is an improvement of \SI{84}{\percent} compared to the cold-started \gls{ocp} which spends approximately three minutes.
The run-time cost of obtaining a feasible trajectory via optimal control is significant compared to performing \astar{} and dynamic generation alone.

\begin{figure}[tb]
	\centering
	\includegraphics[width=0.95\linewidth]{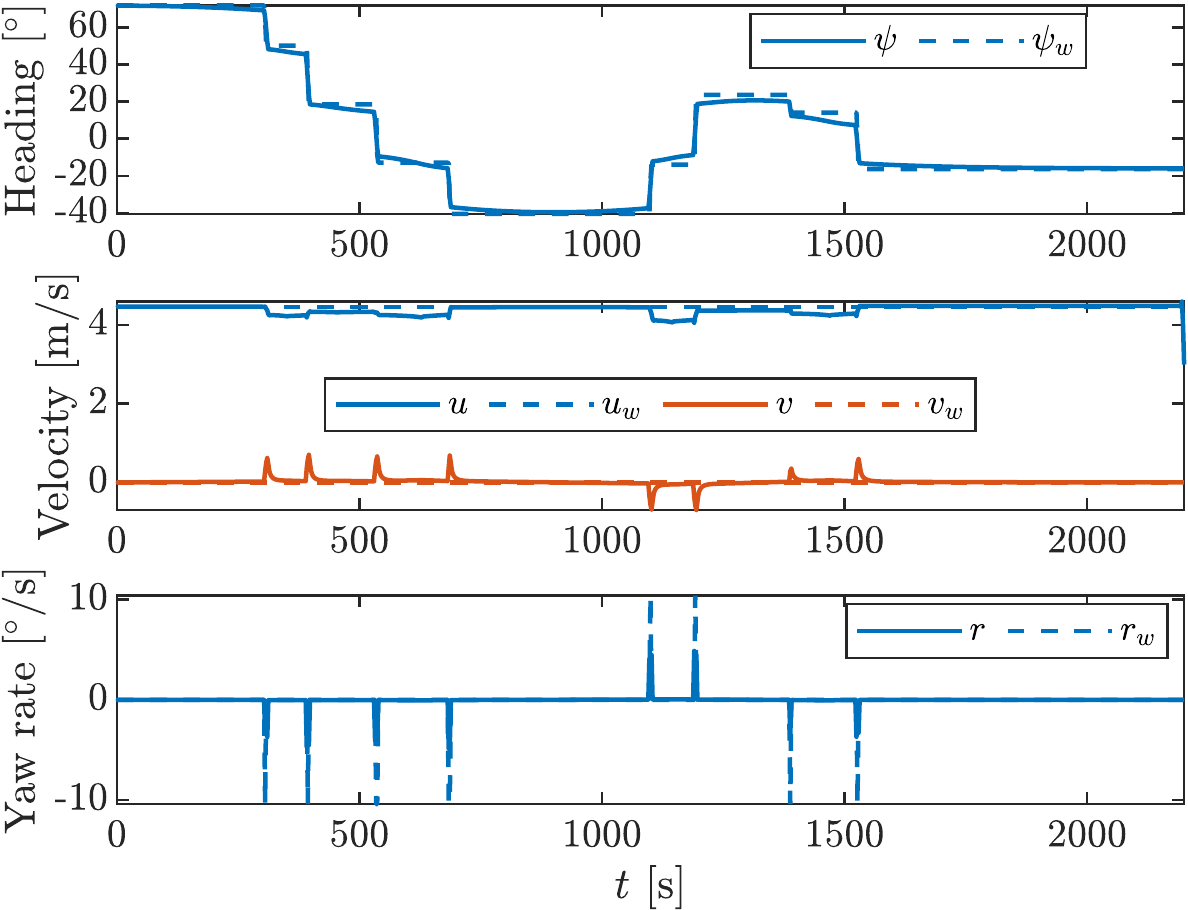}
	\caption{%
		State values for heading, velocities and yaw rate for both the optimized trajectory and the initial guess.
	}
	\label{fig:states}
\end{figure}

\begin{figure}[tb]
	\centering
	\includegraphics[width=0.95\linewidth]{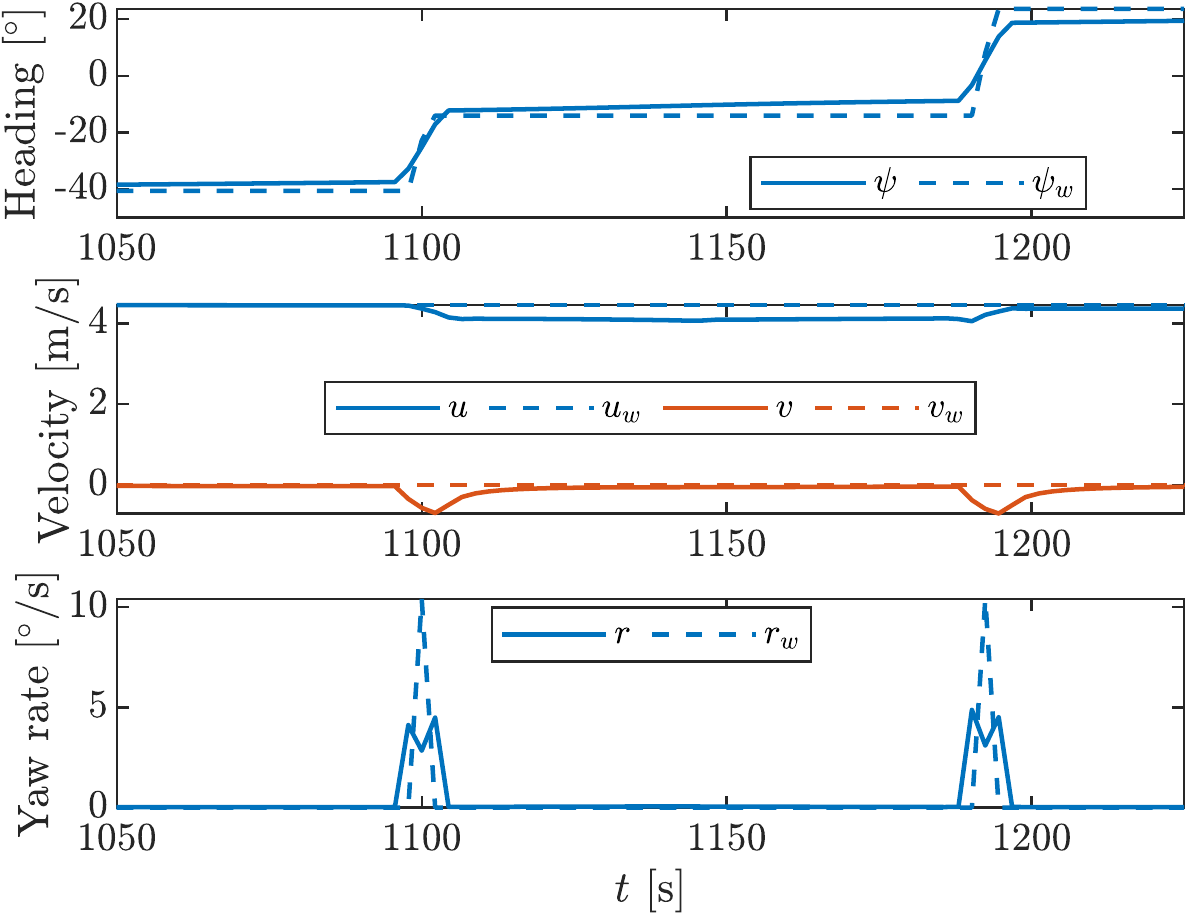}
	\caption{%
		Zoomed-in section of \autoref{fig:states}.
	}
	\label{fig:states-zoomed}
\end{figure}

The state trajectories for the initial guess and warm-started \gls{ocp} are shown in \autoref{fig:states}.
From the heading angle plot, we see that $\psi$ performs jumps of more than \SI{30}{\degree}, which is a clear indication of intent to other vessels, even in situations with restricted visibility \citep{Cockcroft2004}.
This is further observed in the yaw rate state $r$, where instead of having long turns with low yaw rate magnitude, we have abrupt turns with high-valued $r$.
This is shown more clearly in \autoref{fig:states-zoomed}, which zooms in on a selected time interval.

\begin{table}[tb]
	\centering
	\caption{Scenario results.}
	\ifdef{\isarxiv}
	{
	\begin{tabulary}{1.0\linewidth}{LRRR}
	}{
	\begin{tabulary}{1.0\linewidth}{>{\raggedright\hspace{0pt}}m{2.2cm}>{\hspace{-25pt}}R>{\hspace{-25pt}}RR}
	}
		\toprule
											& Warm started \autoref{step:ocp}	& Cold started \autoref{step:ocp} 	& \autoref{step:trajectory} \\
		\midrule
		Feasible 							& Yes								& Yes 								& No \\
		Scaled total cost ($J$) 			& \bf\SI{1.08e4}{}					& \SI{1.54e4}{} 					& \SI{1.13e4}{} \\
		Unscaled energy cost ($J_e$)		& \bf\SI{2.74e7}{[\joule]}			& \SI{3.94e7}{[\joule]} 			& \SI{2.84e7}{[\joule]} \\
		Total run time 						& \SI{26.7}{[\second]}				& \SI{174}{[\second]} 				& \bf\SI{5.7}{[\second]} \\
		\autoref{step:astar} run time 		& \SI{3.4}{[\second]}				& - 								& \SI{3.4}{[\second]} \\
		\autoref{step:trajectory} run time 	& \SI{2.2}{[\second]}				& - 								& \SI{2.2}{[\second]} \\
		\autoref{step:ocp} run time 		& \SI{21.1}{[\second]}				& \SI{174}{[\second]} 				& - \\
		\autoref{step:ocp} iterations		& \bf58								& 549								& - \\
		\bottomrule
	\end{tabulary}
	\label{tab:results}
\end{table}

\section{Conclusion} 
\label{sec:conclusion}

We have developed and demonstrated a pipelined trajectory planning algorithm that exploits the speed and global properties of an \astar{} search with the optimality of an \gls{ocp} solver.
The results from \autoref{sec:simulation_scenario_and_results} show that using the initial guess provided by a smoothed \astar{} path in an \gls{ocp} significantly improves both run time and optimality compared to a cold-started \gls{ocp} alone.
Performing optimization on the \astar{} path significantly increases run time but will find a feasible locally optimal trajectory, as opposed to \astar{} alone.

Qualitatively, the developed method is complete in terms of the \emph{shortest path}, since this is the geometric objective of the \astar{} implementation.
This is dependent on the discretization of the map, since using larger grid spacing to reduce run time removes narrow passages from the solution space.
Using a different discretization scheme such as e.g.\ Voronoi diagrams may guarantee a complete solution space.
The developed method is also locally optimal in the sense of the provided objective, which is a combination of energy consumption and readily observable maneuvers in our case.
The optimality is provided by the implemented \gls{ocp} which alone is not able to find the global optimum, demonstrated by the cold-started result in \autoref{fig:overlaid-paths}.
However, the \gls{ocp} warm-started by the shortest path found by the \astar{} method is at least locally optimal and may be close to the global optimum, since, in the absence of disturbances, the shortest path is also the one that requires the least energy.
In addition to improving optimality of the \astar{} result, the \gls{ocp} adds feasibility, unlike the \astar{} consideration which is purely geometric.
Using this warm-starting scheme is that the \gls{ocp} will lock into one routing alternative.
Depending on the desired sense of optimality, this may not be the desired solution, which is a disadvantage to some use cases.

The algorithm presented here has been used in a hybrid collision avoidance architecture in \citep{Eriksen2019ColregsArxiv}, where it is extended to include disturbances in the form of ocean currents.

Further work on this topic includes:
\begin{itemize}
	\item Implementing a more general obstacle representation to handle a wider range of map representations. 
	E.g.\ the obstacle representation in \citep{Zhang2018} handles convex polygons as smooth inequality conditions.
	\item Improvements on the map discretization scheme are also desirable to reduce computational time of the \astar{} algorithm while preserving completeness of the solution space.
	\item Additionally, an \gls{ocp} representation that is paramet\-riz\-ed by straight lines between waypoints in combination with full-state dynamics may be advantageous to inherently produce \gls{colregs}-compliant trajectories.
\end{itemize}


\section*{Acknowledgements}
	This work is funded by the Research Council of Norway and Innovation Norway with project number 269116.
The work is also supported by the Centres of Excellence funding scheme with project number 223254.


\bibliography{ms}

\end{document}